\documentstyle[preprint,aps]{revtex}

\def\be{\begin{equation}}
\def\en{\end{equation}}
\def\bq{\begin{eqnarray}}
\def\eq{\end{eqnarray}}

\begin{document}

\begin{center}
{\Large {\bf Can the Red Shift be a consequence of the Dilaton Field?}} \\[%
1.5cm]
\end{center}


\begin{center}
{\large {\bf Alejandro Cabo and Alexis Am\'{e}zaga}}

{\it Grupo de F\'{i}sica Te\'{o}rica }

{\it Instituto de Cibernetica, Matem\'{a}tica y F\'{i}sica}\\[0pt]
{\it Calle E No 309, Vedado,La Habana, Cuba }\\[0pt]
\vspace{0.5cm}


{\bf Abstact}
\end{center}

\noindent

The possibility that the expansion rate of the Universe, as reflected by the 

Red Shift, could be produced by the existence of the dilaton field is
explored.

The analysis starts from previously studied solutions of the Einstein
equations

for gravity interacting with a massive scalar field. It is firstly
underlined 

that such solutions can produce the observed values of the Hubble constant.

Since the Einstein-Klein-Gordon lagrangian could be expected to appear as an 

effective one for the dilaton in some approximation, the mentioned solutions

are applied to study this field. Therefore, the vacuum expectation value 

for the dilaton is selected to be of the order of the Planck mass, as it is

frequently fixed in string phenomenology. Then, it follows that the value 

of its effective mass should be as low as m=3.9 10\symbol{94}(-29) cm\symbol{%
94}(-1) in order 

produce the observed expansion rate. The discussion can also predict a
radius of the Universe of the order of 10\symbol{94}(29) cm. Finally, after
adopting the view advanced ina previous work, in which these mentioned
solutions are associated to interior configurations of collapsed scalar
fields, a picture of our Universe as a black hole interior is suggested. 

\setlength{\baselineskip}{1\baselineskip}


\bigskip

\section{Introduction}

The relevance for the physical world of the string theory is an central
issue to be decided for the future research activity in this basic area of
modern theoretical physics [1,2,3]. At present the experimental limitations
make unclear the perspectives for the clarification of this question [4,5].
Therefore, it becomes of central interest the search for any signal in
Nature from the string structure of matter.

As a preamble it could be remarked the construction of realistic models in
four dimensions from superstring theory, the breakdown of various symmetries
in the low energy limit play a very crucial role. At these low energies, it
is commonly expected that a field theoretical effect should be dominant in
order to generate the hierarchy of scales currently observed. In one of the
preferred views: the two steps scenario, string effects dominate to lift
vacuum degeneracy and field theory effects are responsible for to break
supersymmetry. In this approach, the vacuum expectation values (vev) of the
dilaton and other vacuum fields are fixed at high energies with values near
to Planck mass. A problem here resides in the implementation of this
scenario due to no yet complete understanding of nonperturbative string
effects. The search of a deeper knowledge about these effects is currently
very active due to the findings of equivalence of the various string
theories under duality [6].

In this note, our purpose is to explore the possibility for the existence of
a macroscopic signal form the string structure of matter. it would arise
from the relevant scalar field arising in string theory: the dilaton, which
reflects the condensation of zero mass strings [1]. The idea is the
following: As it was recalled above, for the construction of string
phenomenology, the mean vev of the dilaton is fixed at values of the order
of the Planck mass [4,5]. Then we may consider the properties of the dilaton
as input parameters of a particular type of solution of the Einstein
equations for massive scalar field interacting with gravity [7,8]. Such
solutions are regular at a centre of spherical symmetry and show a
decreasing gravitational potential away from the center. The decaying
potential could be interpreted as reflecting the expansion rate of the
Universe. Consider also the possibility that a small value of the mass for
the dilaton can be generated due to combined non-perturbative string
corrections and the breaking of supersymmetry. Then, it seems a sensible
issue to to determine the value of such a mass which can predict the
observed Hubble constant when the dilaton vev is taken of the order of the
Planck mass. The resulting mass values calculated here turn to be extremely
small. This outcome seems at least compatible with the massles of the
dilaton field in the perturbative string theory[1]. Such mass values, up to
our knowledge, could not be rejected under the basis of the current
understanding of string phenomenology.

It should be recognized that the picture here examined would introduce
radical changes in the most accepted views at present about the dynamical
evolution of the Universe [9]. Therefore, a further inspection of its
implication is in need.

It should be also said that the above mentioned solutions corresponds to the
internal part of the extended solutions investigated in [7]. In that work
the central aim was to obtain indications about tht possibility of
interpreting such field configurations as internal space times of black
holes.The external metric in the proposed solutions is the Scwartzschild one
with a null scalar field. An important circumstance in such an
interpretation is that the proper mass of the interior scalar field is
exactly coinciding with the proper mass of a Scwartzschild solution having
the horizon at the precise radius in which the internal metric become
singular [7]. Work is being directed to rigorously argument that these
global field configurations solve the Einstein-Klein-Gordon equation in some
generalized mathematical sense. The complete validation of these
considerations then, would indicate the interpretation of our Universe as an
internal space time of a black hole of collapsed dilaton matter.

\section{1. The special solutions}

The coupled set of Klein-Gordon and Einstein equations in spherical
coordinates which was considered in Ref. [7] has the explicit form

\begin{eqnarray}
{\frac{{u^{\prime }(\rho )}}\rho }-{\frac{{1-u(\rho )}}{\rho {^2}}} &=&{}-{(}%
8\pi k/c^4)({\frac{{u(\rho )\,}\varphi {{{^{\prime }(\rho )}^2}}}2}+{\frac{%
{\it (mc/\hbar )}^2\ \varphi {{{(\rho )}^2}}}2)} \\
{\frac{u(\rho )}{v(\rho )}}\,{\frac{{v^{\prime }(\rho )}}\rho }-{\frac{{%
1-u(\rho )}}{\rho {^2}}} &=&{}+{(}8\pi k/c^4)({\frac{{u(\rho )\,}\varphi {{{%
^{\prime }(\rho )}^2}}}2}-{\frac{{\it (mc/\hbar )}^2{\it \ }\varphi {{{(\rho
)}^2}}}2)} \\
m^2\varphi (\rho )-u(\rho )\,\varphi ^{\prime \prime }(\rho ) &=&\varphi
^{\prime }(\rho )\,\left( {\frac{(u(\rho )+1)}\rho }-{(}8\pi k/c^4){\frac{%
\rho {\it (mc/\hbar )}^2\varphi {(\rho )}^2}2}\right)
\end{eqnarray}

\noindent 
where the invariant interval has been taken as defined by the spherical
coordinates form

\begin{eqnarray*}
ds^2 &=&{\it v}(\rho ){dx^o}^2-u(\rho )^{-1}d\rho ^2-\rho ^2(sin^2\theta
d\phi ^2+d\theta ^2). \\
{dx^o} &=&cdt
\end{eqnarray*}

The scalar field $\phi $ is real and its mass $m$ can be absorbed in the
definition of a new radial variable $r=mc\rho /\hbar $. Moreover, the scalar
field also can been scaled as $\varphi =\phi /\sqrt{8\pi k/c^4\ }$in order
to absorb the factor $8\pi k/c^4$ multiplying the energy momentum tensor in
it. After these changes, the equations take the form

\begin{eqnarray}
{\frac{{u^{\prime }(r)}}r}-{\frac{{1-u(r)}}{{r^2}}} &=&{}-{\frac{{u(r)\,\phi 
{{^{\prime }(r)}^2}}}2}-{\frac{\phi {{{(r)}^2}}}2} \\
{\frac{u(r)}{v(r)}}\,{\frac{{v^{\prime }(r)}}r}-{\frac{{1-u(r)}}{{r^2}}}
&=&{}+{\frac{{u(r)\,\phi {{^{\prime }(r)}^2}}}2}-{\frac{\phi {{{(r)}^2}}}2}
\\
\phi (r)-u(r)\,\phi ^{\prime \prime }(r) &=&+\phi ^{\prime }(r)\,\left( {%
\frac{(u(r)+1)}{r\ }}-{\frac{r\phi {(r)}^2}{2\ }}\right)
\end{eqnarray}

It can be observed that equations (4) and (6) are closed because they do not
depend on v$(r)$. In place of v(r)$,$ below it will be sometimes used the
variable

\[
\nu (r)=\log \,(\text{v}(r)). 
\]

The set of equations (4)-(6) have solutions showing a leading asymptotic
behavior near the origin $r=0$ of the following form

\begin{eqnarray}
u(r) &=&1-{\frac{{{{\it \phi _0}^2}\,{r^2}}}6}+... \\
\text{v}(r) &=&1-{\frac{{{{\it \phi _0}^2}\,{r^2}}}6}+... \\
\phi (r) &=&{\it \phi _0}\,\left( 1+{\frac{{r^2}}6}\right) +...
\end{eqnarray}

The behavior of this solutions motivated the present note. As it can be
noticed from them, the gravitation potential decreases away from the origin
of coordinates with a quadratic dependence with the radial distance. Near
the origin, the metric is approximately lorentzian. Then, some questions are
directly suggested : Could these solutions be applied to construct a model
of the expansion rate of the Universe?, assuming that so is the case: What
is the nature of the scalar field being considered?. Similar questions have
been addressed in the literature in connection with the possible role of a
cosmological constant in Robertson-Walker (WR) cosmologies [9] . The
existence of physical sources of such cosmological constants has been a
limitation for the introduction of a cosmological term in such models (It
may be useful to say that the massive term for the scalar field plays a
similar role in the dynamical equations, that a cosmological constant if the
field is space-time independent in some region). However, at present there
are new theoretical ideas related with string theory phenomenology which
assume the existence of a vacuum expectations of scalar fields. These values
could be suspected to be connected the parameter ${\it \phi _0}$ in
equations (7)-(9 ). Concretely, as it was mentioned before, in one of the
adopted views about the construction of string phenomenological theories, a
high value of the vev for the dilaton field of the order of the Planck mass
is fixed [4,5]. On another hand, the perturbative value of the dilaton mas
is null, a fact that can cast doubts on the present proposal. However, it
can be taken into account the possibility that a mass can be generated by
string non perturbative corrections and also by the various spontaneous
symmetry breaking effects that are expected to occur in order to reproduce
the observable experience. Therefore, it seems reasonable to assume non
vanishing values for the dilaton effective mass. As it will be seen, the
magnitude of the of the product of the dilaton mass and its vev is
determined from the condition of fixing the observable value of the Hubble
constant.

Let us consider the equation (8) which defines the gravitation potential in
the vecinity of the origin when written in the form 
\begin{equation}
\text{v}(\rho )=1+2\Phi (\rho )/c^2,
\end{equation}

where $\Phi (\rho)$ is the gravitational potential.

Then, for $\Phi $ it follows 
\begin{equation}
\Phi (\rho )=-(4\pi /6)G_N\phi _0^2(m/h)^2\rho ^2.  \label{II4}
\end{equation}
The Hubble law is now expressed by the relation

\begin{eqnarray}
E_{Pot}(0) &=&0=E_{Pot}(\rho )+E_{Kin}(\rho )  \nonumber \\
&=&M\ \Phi (\rho )+M\frac{V^2}2  \label{II5}
\end{eqnarray}

where the l.h.s. is the potential energy of a gallaxy of mass M at the
origin of coordinates and V$(\rho )=d\rho /dt$ is its velocity at the radial
distance $\rho .$

Recalling the definition of the Hubble constant $H_0$ through $(V/\rho
)^2=H_0^2$ the following relation between the mass and the vev of the scalar
field follows 
\begin{equation}
m^2\phi _0^2=\frac{\hbar ^2H_0^2}{4\pi k}  \label{II6}
\end{equation}

After assuming the values for $H_0=75\times 10^5\hspace{4mm}cm/(sMpc)$ and
the gravitation constant $k=6.67\times 10^{-8}\hspace{4mm}cm/gs^2$ the
product of the mass and the vev takes the value (in rationalized units: $%
[m]=cm^{-1}$, $[\phi ]=cm^{-1}$)

\begin{equation}
m\phi _0=24584\hspace{4mm}cm^{-2}.  \label{II7}
\end{equation}

The relationship (\ref{II7}) has been gotten starting from a solution for a
massive scalar field in interaction with the gravity, within the context of
general classical relativity. However, as it was discussed in Ref.[8] the
Einstein Klein Gordon equations can be expected to become a reasonable
effective action for the dilaton field in a consistent context of gravity
based in superstring theory, if the dilaton field aquires mass through any
mechanism.

Finally after setting the vev of the dilaton to 1 in the Planck scale, i.e.

$<\phi _0>=0.63\ 10^{33}cm^{-1}$ using (\ref{II7}) it is obtained

\begin{equation}
m=3.9\ 10^{-29}cm^{-1}..  \label{III1}
\end{equation}

Therefore, if the nonperturbative corrections or symmetry breaking effects
are able to produce a tiny non vanishing dilaton mass, it would be
sufficient to generate the observed expansion rate of the gallaxies.

\medskip

Let us consider now an additional implication of the possibility being
discussed.

As it was mentioned in the introduction, the approximate solutions near $r=0$
can be numerically extended away from the origin by taking as the initial
conditions the values at small radial distance of $u(r),$v$(r),\phi (r)$ and 
$\phi ^{\prime }(r)$. The value of the vev is taken as $\phi _0=4.5$ in
(7)-(9) which is determined for a fixed magnitude of the dilaton vev of the
order of the Planck mass. The initial values and the first derivative of the
scalar field necessary for the numerical algorithm were calculated also from
(7)-(9) at the radial position $r=0.003$.

The results of [7] show a decreasing behavior of the function $u(r),$ which
approaches zero linearly at some radius $r_0=0.502416$. The picture for the
variation of v$(r)$, while similarly decreasing, tends to approach a
constant value differing from zero when $r$ approaches $r_0$. The scalar
field, on another hand, increases away from the origin and approaches a
constant value at $r_0$, but with a fast growing slope diverging at $r_0$.


The asymptotic behavior near $r_0$ of the numerical solution was also
determined in [7] to be given by

\begin{eqnarray}
u(r) &=&(r_0\phi _{r_0}^2-2/r_0)(r_0-r)+...  \label{II2} \\
\phi (r) &=&\phi _{r_0}-2\sqrt{r_0-r}/{\sqrt{r_0}}+... \\
&&
\end{eqnarray}

It was shown in [7] that the singular behavior near $r_0$ of the numerical

solution is well reproduced by this asymptotic expressions. An important
property

of the dependence with the radial distance of u(r) is the fact that this
quantity tends

to vanish at $r=r_0$. This fact allows to show that the proper mass of the
scalar field

in the interior region exactly coincides with the proper mass of a
Scwartzschild solution having its horizon at the same radius $r_0$. This
circumstance strongly suggest the interest of a global field configuration
defined by the here considered one at the internal region after extedending
it to coincide witht the Scwartzschild solution at exterior by also taking
null values of the scalar field (as suggestd by the non-hair theorems). Such
fields evidently solve the dynamical equations everywhere except at of the
shell $r=r_0$. Then, if some localized sources axist at the boundary, they
should not contribute to the total value of thee proper mass. This remarks
strongly indicate that such fields should correspond to solutions in all the
space in some generalized mathematical sense. Work on the verification of
this possibility is under developement.

\section{ Conclusions}

An alternative for the physical interpretation of the observed expansion
rate of the Universe is explored. It rests in the assumption that the
dilaton field in string theory gets a small mass under non-perturbative
corrections or the symmetry breakings needed for string theory to become a
realsitic theory. The validity of the picture would implicate the
interpretation of the Universe as the interior region of a black hole formed
in the collapse of dilaton field matter. It is clear that the formulation of
the proposed model should undergone a closer scrutiny in connection with its
ability to explain the many observed cosmological data. The inevestigation
on these questions would be considered elsewhere.

The discussion has some resemblance with the analysis of Robertson-Walker
models under assuming a non-vanishing cosmological constant [9]. It could be
imagined that what is done here is to explore the possibility that the vev
of the dilaton field play a similar role that a cosmological constant in
determining the expansion rate [9]. \newpage

\noindent {\large {\bf References}}

[1] M. Green, J. Schwartz and E. Witten,{\it \ }${\it SuperstringTheory,1,2,}
$Cambridge

University Press(1987)

[2] M. Kaku, {\it String Theory}, Springer Verlag (1998).

[3] J. Polshinski, {\it Les Houches Lectures, hep-th/9411028 (1994).}

[4] F. Quevedo, {\it Superstring Phenomenoloy, An Overview}, hep-ph/9707434
(1997).{\it \ }

[5] A. Font, L.E. Ib\'{a}\~{n}ez, F. Quevedo and A. Sierra, {\it The
Construction of }

{\it ''Realistic'' Four Dimensional Strings Through Orbifolds,
CERN-TH-5326/89.}

[6] R. Diskgraaf, {\it Les Houches Lectures on Fields, Strings and Duality},

hep-th/9703136 (1997).

[7] A. Cabo and E. Ay\'{o}n, accepted in Int. Jour. Mod. Phys. A,
gr-qc/9704073

(1997).

[8] M. Fabbrichessi and R. Iengo, Phys.Lett. B292 (1992)262, hep-th/9207087

(1992).

[9] R. Schaefer,Page 649, in:{\it \ ICTP Series in Theoretical Physics, Vol
9, 1992 }

{\it Summer Workshop in High Energy Physics and Cosmology,} World

Scientific,1992.

%
end{document}

\end{document}